\documentclass[aps,twocolumn,prl,preprintnumbers,showpacs,amsmath,amssymb,superscriptaddress]{revtex4-1}
\usepackage{graphicx}
\usepackage{dcolumn}
\usepackage{threeparttable}
\usepackage{multirow}
\usepackage{booktabs}
\usepackage{txfonts}
\usepackage{xcolor}
\usepackage{bm}
\usepackage{amssymb}
\usepackage{amsmath}
\usepackage{latexsym}
\usepackage{epsfig}
\usepackage{amsbsy}
\usepackage{array}
\usepackage{tabularx}

\begin{document}

\title {Anisotropic Magnetic Couplings and Structure-Driven Canted to Collinear Transitions in Spin-orbit
Coupled Sr$_2$IrO$_4$}

\author{Peitao Liu}
\affiliation{Faculty of Physics, Computational Materials Physics, University of Vienna, Vienna A-1090, Austria}
\affiliation{Shenyang National Laboratory for Materials Science, Institute of Metal Research, Chinese Academy of Sciences, Shenyang 110016, China}

\author{Sergii Khmelevskyi}
\affiliation{Faculty of Physics, Computational Materials Physics, University of Vienna, Vienna A-1090, Austria}
\affiliation{Department of Theoretical Physics, Budapest University of Technology and Economics, H-1111, Budapest, Hungary}

\author{Bongjae Kim}
\affiliation{Faculty of Physics, Computational Materials Physics, University of Vienna, Vienna A-1090, Austria}

\author{Martijn Marsman}
\affiliation{Faculty of Physics, Computational Materials Physics, University of Vienna, Vienna A-1090, Austria}

\author{Dianzhong Li}
\affiliation{Shenyang National Laboratory for Materials Science, Institute of Metal Research, Chinese Academy of Sciences, Shenyang 110016, China}

\author{Xing-Qiu Chen}
\affiliation{Shenyang National Laboratory for Materials Science, Institute of Metal Research, Chinese Academy of Sciences, Shenyang 110016, China}

\author{D.D. Sarma}
\affiliation{Solid State and Structural Chemistry Unit, Indian Institute of Science, Bangalore-560012, India}

\author{Georg Kresse}
\affiliation{Faculty of Physics, Computational Materials Physics, University of Vienna, Vienna A-1090, Austria}

\author{Cesare Franchini}
\email[Corresponding author: ]{cesare.franchini@univie.ac.at}
\affiliation{Faculty of Physics, Computational Materials Physics, University of Vienna, Vienna A-1090, Austria}

\begin{abstract}

We put forward a scheme to study the anisotropic magnetic couplings in Sr$_2$IrO$_4$ by mapping fully relativistic
constrained noncollinear density functional theory including an on-site Hubbard $U$ correction onto a general
spin model Hamiltonian.
This procedure allows for the simultaneous account and direct control of the lattice, spin and orbital interactions within a fully
\emph{ab initio} scheme. We compute the isotropic, single site anisotropy and Dzyaloshinskii-Moriya (DM) coupling
parameters, and clarify that the origin of the canted magnetic state in Sr$_2$IrO$_4$ arises from the interplay
between structural distortions and the competition between isotropic exchange and DM interactions.
A complete magnetic phase diagram with respect to the tetragonal distortion and the
rotation of IrO$_6$ octahedra is constructed, revealing the presence of two types of canted to collinear magnetic
transitions: a spin-flop transition with increasing tetragonal distortion and a complete quenching of the
basal weak ferromagnetic moment below a critical octahedral rotation.
\end{abstract}

\maketitle

In weak ferromagnetic materials the subtle interplay among different types of magnetic interactions can cause
the formation of complex canted spin structures involving the so-called Dzyaloshinskii-Moriya (DM) effect,
arising from the coupling between the spin and orbital angular momenta \cite{Dziailoshinskii57, Moriya1960}.
Intense research was done in this field in the last few years, motivated by the foreseeable applications
in storage technology and by the air of mystery enveloping the quantum-mechanical origin of DM structures
\cite{Dmitrenko2014, Khaliullin2009, Bode2007, Sergienko2006}.
A crucial aspect of the DM systems is the entanglement between structural distortions and magnetism,
which could be exploited as a way to tune the spin texture by modifying the structure upon external
stimuli such as pressure and strain \cite{Khaliullin2009, Haskel2012, Lupascu2014}.

\begin{figure}[h!]
\begin{center}
\includegraphics[width=0.45\textwidth]{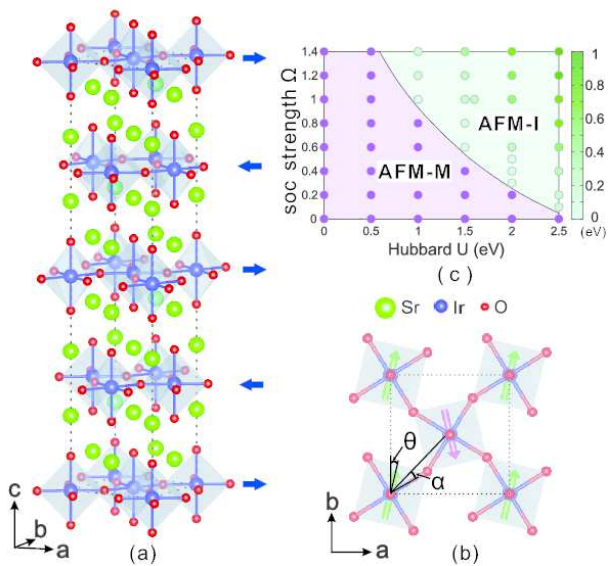}
\end{center}
\caption{(color online) (a) The unit cell of Sr$_2$IrO$_4$ containing 56 atoms.
(b) Projection onto the $ab$ plane
showing the octahedral tilting angle $\alpha$ and the canting angle ${\theta}$
(c)  Metal (M) to insulator (I) transition phase diagram
as a function of the Hubbard $U$ and the SOC strength ${\Omega}$. Filled circles are the computed
($U$, ${\Omega}$) points. The right (coloured) bar indicate the value of the gap (in eV) in the
insulating regime.}
\label{fig:01}
\end{figure}

The cross coupling between the different electronic, lattice and spin degrees of freedom
is particulary rich in iridates. Here, the spin-orbit coupling (SOC), electron-electron correlations,
and spin-exchange interactions operate with comparable strengths and gives rise to a large variety
of exotic states \cite{Shitade2009,Kim2008,Moon2008, Kim2009, Chaloupka2010, Pesin2010, Okada2013}.
The most striking example of this class of materials is the layered perovskite Sr$_2$IrO$_4$,
characterized by a novel relativistic Mott insulating state  \cite{Kim2008, Kim2009, Moon2008, Moon2009, Watanabe2010}
and an unusual in-plane canted antiferromagnetism (AFM) with a weak net ferromagnetic (FM)
component  \cite{FujiyamaPRL2012, Khaliullin2009}. The small electronic gap ($\approx 0.3$ eV \cite{Moon2009})
is opened by modest Hubbard interactions ($U$ $\approx$ 1.5-2 eV  \cite{Arita2012})
and by the strong spin-orbit coupling (${\xi}_{soc}\approx 0.5$ eV  \cite{Katukuri2012})
which effectively narrows the $d$ orbital bandwidth and give rise to an ideal
$J_\text{eff}$=1/2-like state  \cite{Kim2008, Zhang2013,Fujiyama2014}. This is considered to be robust despite
the presence of noncubic structural distortions  \cite{Boseggia2013PRL}.
Neutron diffraction experiments indicate that the IrO$_6$ octahedra
are rotated by $\alpha =$11.5$^{\circ}$  and elongated in the $c$ direction
($c/a \approx$ 1.04) \cite{Crawford1994}, generating the enlarged $\sqrt{2}a\times 2c$ $I4_1/acd$
tetragonal cell shown in Fig. \ref{fig:01}(a). The spins, coupled with the orbital
moment, follow the rotation of the octahedra and establish an insulating canted AFM ordering
with a canting angle ${\theta} \approx 12.2^{\circ}$
\cite{Ye2013, Boseggia2013} [Figs. \ref{fig:01}(a) and \ref{fig:01}(b)].
The formation of the canted AFM ordering is believed to arise from the entanglement of structural
distortions and SOC, which affects the balance between the DM anisotropic interaction and
the isotropic magnetic coupling. A simultaneous treatment of all -- lattice, spin and orbital --
effects is therefore essential for an accurate account of magnetic interactions in this type of
magnetically canted structures.

The aim of our study is to propose a general scheme to treat and explain canted magnetic states
in spin-orbit coupled systems fully \emph{ab initio}. We do this by introducing a series
of directionally constrained non-collinear spin setups and self-consistent total energy
calculations. We determine the magnitude of all relevant magnetic interactions in Sr$_2$IrO$_4$,
including the isotropic exchange, the single-ion anisotropy and the DM coupling, and study the
influence of structural perturbations (tetragonal distortion and octahedral rotation) on
the spin ordering.

The theoretical modelling of canted states can be conducted by model Hamiltonian or first-principles approaches.
Based on a model Hamiltonian approach Jackeli and Khaliullin determined that magnetic interactions
in Sr$_2$IrO$_4$ are governed by the lattice geometry enabling the possibility to tune the spin ordering
by small structural perturbations. For instance, they found that a spin-flop transition to $z$-collinear
AFM ordering can be obtained above a critical tetragonal distortion strength \cite{Khaliullin2009}.
The calculation of the relative strength between the isotropic and anisotropic interactions is theoretically
and computationally challenging. While the isotropic contributions can be estimated accurately from first-principles
within density functional theory (DFT), beyond-DFT methods and multiple scattering
theory \cite{Archer2011,Dane2008}, the accurate estimation of the DM coupling is extremely difficult.
Using a phenomenological microscopic model with a set of optimally chosen parameters Kim and colleagues
found  a large DM coupling parameter $D$ = 26.2 meV, fingerprint of a strong SOC, and a
$\lvert D \rvert$ / $\lvert J \rvert$ ratio of 0.34, yielding $\theta \approx 9.3^{\circ}$ in good
agreement with experiment \cite{Kim2012}.
The \emph{ab inito} calculation of $D$ for periodic systems can be achieved by perturbation theory
and Green's function technique \cite{Solovyev1996, Dmitrenko2014}, scattering theory \cite{Santos2011},
or within constrained non-collinear spin DFT \cite{Kubler200, Hobbs2000}, through the spin-orbit induced
corrections to the energy of spiral spin-density waves \cite{Heide2009}.

Here we follow an alternative route based on mapping self-consistent DFT total energies of several
non-collinear magnetic configurations, obtained by constraining the magnetic moment along
specific directions, onto a general spin-dependent Hamiltonian. In our approach, we consider the cross-coupling
between spin, structural relaxations and spin-orbit interaction fully self consistently and within the same
theoretical framework.
We clarify that the origin of the canted AFM state in Sr$_2$IrO$_4$ arises naturally from the fine balance between the
isotropic ($JS^2$= -0.32 meV), and DM ($D_zS^2$= -0.25 meV) couplings, and construct a detailed magnetic phase diagram showing the transition
from the the canted ground state to different types of collinear magnetic states as a function of the tetragonal
distortion  $c/a$ and the octahedral rotation angle $\alpha$ entirely \emph{ab initio}.

We use the Vienna \emph{Ab initio} Simulation Package (VASP) \cite{Kresse-1, Kresse-2} and adopt a fully relativistic
(i.e., including SOC) DFT scheme within the gradient corrected approximation (GGA) with the inclusion of an on-site Hubbard
$U$ correction \cite{Dudarev2} to account for electron screened exchange in the Ir $d$ manifold.
The reciprocal space is sampled with a 7$\times$7$\times$3 $\textbf{k}$-mesh and energy cut off up to 800 eV with a
high convergence precision of 10$^{-8}$ eV is necessary to achieved well converged results in the meV scale.
The atomic positions are optimized with the lattice parameters fixed at the experimental values \cite{Crawford1994}.

This GGA+$U$+SOC approach correctly describes the canted insulating ground state of Sr$_2$IrO$_4$
in line with previous LDA+$U$+SOC studies  \cite{Kim2008}. The gap is opened by the interwoven
action of SOC and $U$ as shown in the SOC/$U$ metal-insulator phase diagram of Fig.\ref{fig:01}(c),
constructed by tuning the SOC strength through the scaling factor $\Omega$ ($\Omega =1$ correspond to the
self-consistent SOC strength) for values of $U$ ranging from 0 to 2.5 eV.
From the metal-insulator phase diagram it can be seen that neglecting either the SOC ($\Omega$ = 0) or
 $U$ ($U$ = 0) Sr$_2$IrO$_4$ remains metallic: SOC and $U$ alone cannot activate any metal-to-insulator transition
(MIT). The MIT is driven by the cooperative action of SOC and $U$.
As a general tendency we observe that the stronger the SOC strength, the smaller is the critical interaction $U$
required for opening the gap. By fitting the $U$ to the experimental bandgap
($\approx 0.3$ eV \cite{Moon2009}) for $\Omega$ = 1 we obtain an optimal $U$ of 1.6 eV,
similar to the constrained random phase approximation estimation, 1.96 eV \cite{Arita2012}.
From the total energy difference with and without the inclusion of SOC we have estimated a
SOC energy ${\xi}_{soc}$ of 0.7 eV/Ir, in line with electron spin resonance
measurements\cite{Andlauer1976}, which is only about half of the computed value of $U$.
These results provide clear evidence for the relativistic Mott-Hubbard character of the insulating state in
Sr$_2$IrO$_4$. Moreover, the strong ligand field energy 10$Dq$ (3.7 eV, computed
from the splitting between the $e_\text{g}$ and $t_\text{2g}$ orbitals) pushes down the
$J_\text{eff}$=3/2 manifold and promotes a $J_\text{eff}$=1/2 like state compatible with previous
interpretations \cite{Kim2008}.
Our estimated values of $U$, ${\xi}_{soc}$ and 10$Dq$ are collected in Table. \ref{tab:01}, along
with available literature data.

Moreover, our approach successfully reproduces the in-plane canted AFM state with a weak
net moment of about 0.08 ${\mu}_B$/Ir, in good agreement with the experimental values of
0.06$\sim$0.1 ${\mu}_B$/Ir  \cite{Kim2009,FujiyamaPRL2012,Crawford1994}.
The components of the orbital (0.27 ${\mu}_B$/Ir) and spin (0.12 ${\mu}_B$/Ir) moments are antiferromagnetically
aligned along the $b$ axis and ferromagnetically canted with respect to the $a$ axis [see Fig. \ref{fig:01}(b)].
The calculated total canting angle is $\theta$ = 12.3$^{\circ}$
(with similar spin, $\theta_s = 14.4^{\circ}$, and orbital, $\theta_o = 11.2^{\circ}$, components)
is in excellent agreement with the experimental value of 12.2$^{\circ}$  \cite{Boseggia2013}.

After providing a detailed description of the ground state properties of Sr$_2$IrO$_4$, we can now address
and interpret the origin of this exotic anisotropic magnetic structure by an \emph{ab initio} evaluation
of the magnetic coupling  parameters within the framework of constrained DFT+$U$ for noncollinear magnetism.
To this end, we consider a penalty contribution to the total energy which fixes the local moment into a specific direction
\begin{equation}
E= E_0+\sum_{i}\lambda[\mathbf{M}_i-\mathbf{M}^0_i(\mathbf{M}^0_i\cdot\mathbf{M}_i)]^2,
\label{eq0}
\end{equation}
where $E_0$ is the usual DFT energy without any constraint and the second term represents the penalty
energy due to the non-collinear directional constraint. ${\mathbf{M}^0_i}$ is
a unit vector along the desired direction of the magnetic
moment at site $i$ and $\mathbf{M}_i$ is the integrated magnetic moment inside the Wigner-Seitz cell around atom $i$,
whereas the parameter $\lambda$ controls the penalty energy contribution, which becomes vanishingly small by increasing
$\lambda$.
Thus, by progressively increasing $\lambda$, one converges towards the DFT total energy for a given constrained
noncollinear magnetic configuration. We found that $\lambda$=10 is generally sufficient to reduce the
penalty energy to $<$ 10$^{-5}$ eV.

\begin{table}[t!]
\caption{Hubbard $U$ (eV), ligand field energy 10$Dq$ (eV), spin-orbit
coupling energy ${\xi}_{soc}$ (eV), isotropic AFM exchange $JS^2$ (meV), single-ion anisotropy $K$ (meV),
and the DM exchange $D_zS^2$ (meV) of Sr$_2$IrO$_4$. Available literature data are given with respect to the
calculated $S$=0.12 ${\mu}_B$.}
\begin{ruledtabular}
\begin{tabular}{ccccccc}
                           & 10$Dq$ &    $U$     &     ${\xi}_{soc}$     &  $JS^2$ & $K$    &  $D_zS^2$       \\
This work                  & 3.7    &   1.6      &      0.7              & -0.32   &  -0.10 & -0.25        \\
Literature                & 3.8$^a$ &   1.96$^b$ &      0.46$^c$         & -0.73$^d$, -0.86$^e$ & -     & -0.38$^f$          \\
\end{tabular}
\begin{tablenotes}
\item [a]$^a$Reference  \cite{Sala2014-2}
\item [b]$^b$Reference  \cite{Arita2012}
\item [c]$^c$Reference  \cite{Katukuri2012}
\item [d]$^d$Reference  \cite{Katukuri2012}
\item [e]$^e$Reference  \cite{cuprate-1}
\item [f]$^f$Reference  \cite{Kim2012}
\end{tablenotes}
\end{ruledtabular}
\label{tab:01}
\end{table}

\begin{figure}[h!]
\begin{center}
\includegraphics[clip= , width=0.45\textwidth]{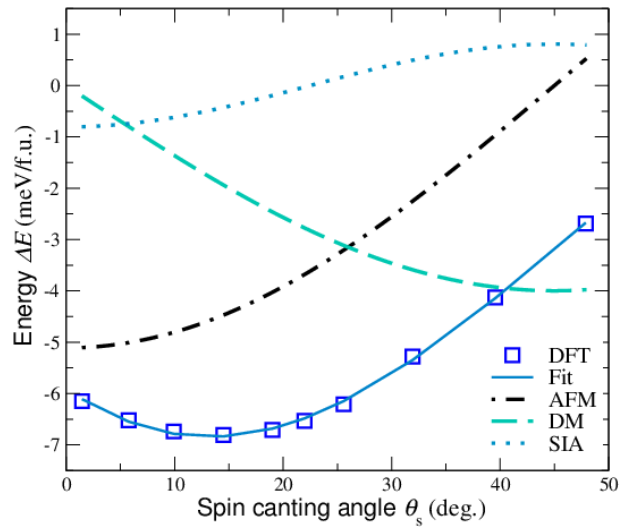}
\end{center}
\caption{(color online) The spin-dependent total energy as a function of the in-plane spin canting angle $\theta_s$
decomposed over the isotropic AFM exchange, SIA and DM contributions.}
\label{fig:02}
\end{figure}

By varying the spin canting angle $\theta_s$ within the $ab$ plane from 0 ($ab$-collinear AFM) to 45$^{\circ}$
(perpendicular alignment of nearest neighbor spin moments) we obtain the DFT total energy curve shown in Fig. \ref{fig:02}.
To calculate the magnetic coupling parameters, we start from the conventional expression for a spin-dependent energy
\begin{equation}
\Delta E= -\sum_{i< j}\textbf{J}_{ij}\textbf{S}_{i}\cdot\textbf{S}_{j}+\sum_{i}\varepsilon^{i}_{an}(\textbf{S}_{i})+\sum_{i<j}\textbf{D}_{ij}\cdot[\textbf{S}_{i}\times\textbf{S}_{j}],
\label{eq1}
\end{equation}
where the first, second and last term represent the isotropic AFM exchange,  the single-ion
anisotropy (SIA) and Dzyaloshinskii-Moriya interaction, respectively.
Considering that Sr$_2$IrO$_4$ adopts a quasi two-dimensional geometry (see Fig.\ref{fig:01}), the interlayer
hopping is much smaller than the in-plane one and can be neglected; we therefore consider the in-plane nearest-neighbor
hopping only.  Also, within the in-plane canted AFM state the DM anisotropy has the $z$ component only.
In addition, the single-ion anisotropy reduces to $\varepsilon_{an}(S)=K\cos(4\theta{_s})$ as a result of the
tetragonal symmetry in the lattice. After taking the sum over all the Ir ions in the conventional unit cell,
Eq. (\ref{eq1}) finally reduces to,
\begin{equation}
\Delta E=16JS^2\cos(2\theta_s)+8K\cos(4\theta_s)+16D_{z}S^2\sin(2\theta_s)
\label{eq2}
\end{equation}
where $J$, $K$ and $D_{z}$ are the nearest-neighbor isotropic exchange, SIA
and DM coupling parameters, respectively, and  $S$ is the magnitude of the
in-plane spin moment (see Fig. \ref{fig:01}).

The magnetic coupling parameters $J$, $K$ and $D_{z}$ can be extracted by fitting the first-principles
calculated data using the model Eq. (\ref{eq2}). The obtained results, $JS^2$ = -0.32 meV, $K$ = -0.1 meV, and
$D_zS^2$ = -0.25 meV, collected in Table \ref{tab:01}, 
compare well with phenomenological data available
in literature and elucidate the canted state completely \emph{ab initio}.
The AFM isotropic exchange, favoring a collinear alignment of the spins, is the dominating interaction and
aids the stabilization of the AFM ordering. The DM interaction, which prefers an orthogonal coupling of the spins,
is only slightly weaker than the isotropic exchange and assists the formation of a canted spin arrangement.
The negative sign of $D_z$ indicates that the $\textbf{D}$ vector is antiparallel to
the vector product [$\textbf{S}_{i}\times\textbf{S}_{j}$] in order to reduce the energy of the system.
The SIA is smaller than the isotropic and DM interactions, but still favors in-plane magnetism at small canting angles,
as can be seen from the decomposition of the energy curve into its isotropic, DM, and SIA contributions (Fig. \ref{fig:02}).
As expected, the evolution of the DM and isotropic energies as a function of $\theta_s$ follows a different trend.
The DM term become stronger by increasing $\theta_s$ with a peak at 45$^{\circ}$ (perpendicular spin ordering)
whereas the AFM isotropic exchange is more active for small canting angles and reaches its maximum
strength at the onset of the collinear regime ($\theta_s$ = 0$^{\circ}$). The formation of the canted in-plane state at
$\theta_s$ = 14.4$^{\circ}$ is the results of the subtle competition between these three terms, $JS^2$, $D_zS^2$ and $K$.

Having identified the origin of the anisotropic magnetism, we are now in a position to define
the role of the structural distortions on the magnetism and in particular on the way the
$c/a$ ratio and the IrO$_{6}$ octahedra angle $\alpha$ mediate the spin canting.
In fact, due to the strong SOC, structural distortions inevitably affect the shape of the
$t_{2g}$ orbitals and directly influence the magnetic interactions.

\begin{figure*}
\begin{center}
\includegraphics[width=0.95\textwidth]{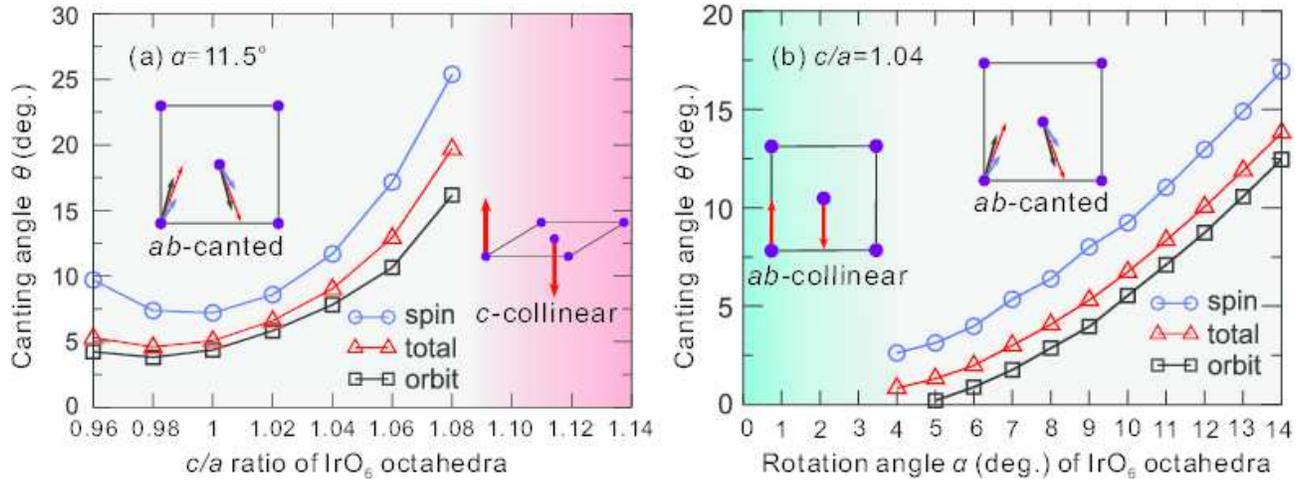}
\end{center}
\caption{(color online) Change of the magnetic canting angle $\theta$ (spin, orbital and total moments)
as a function of the structural distortions. (a) $\theta$ vs. $c/a$ at the optimized octahedra rotation angle
${\alpha}=11.5^{\circ}$; (b) IrO$_6$ angle ${\alpha}$ vs. $\theta$ at the optimized
tetragonal distortion $c/a=1.04$. Shaded areas indicate regions with a different spin ordering
($ab$-canted, $ab$-collinear and $c$-collinear, schematically represented in the insets).}
\label{fig:03}
\end{figure*}

Figure \ref{fig:03} shows the evolution of the canting angle $\theta$ as a function of the tetragonal distortion
(controlled by the octahedral $c/a$) and octahedral rotation angle ${\alpha}$. The data, decomposed
over the spin, orbital and total contributions, are obtained by varying $c/a$ and keeping $\alpha$ fixed to
its optimized value ($11.5^{\circ}$) and vice versa.
The plots show a strong coupling between the lattice and magnetic channels and
very similar curves are found for the spin and orbital part.

For a fixed $\alpha$, the canting angle $\theta$ exhibits a parabolic behaviour as a function $c/a$ [Fig \ref{fig:03}(a)].
At the cubic limit of $c/a=1$, $\theta$ ($\approx$ 5.1$^{\circ}$) does not vanish, indicating that the strength of the
DM coupling is strong enough to produce magnetic canting in the presence of a finite octahedral rotation angle.
For larger tetragonal distortions $c/a > 1.09$,
a spin-flop non-collinear to $c$-collinear transition occurs, supporting the microscopic predictions by Jackeli and
Khaliullin  \cite{Khaliullin2009}.

When the $c/a$ is fixed [Fig. \ref{fig:03}(b)], the magnetic canting angle $\theta$ follows
the octahedral rotation angle $\alpha$ due to the strong SOC, confirming the robustness of the
$J_\text{eff}$=1/2-like state even in the presence of a tetragonal distortion, in agreement with
experimental observations  \cite{Boseggia2013}. For $\alpha$ smaller than the critical
value of 4 $^{\circ}$, the weak ferromagnetic moment is quenched and Sr$_2$IrO$_4$ converts to an in-plane
collinear AFM insulator, consistent with the pressure-induced canted-to-collinear transition
inferred from the high-pressure experiments \cite{Haskel2012}.

\begin{figure}
\begin{center}
\includegraphics[width=0.45\textwidth]{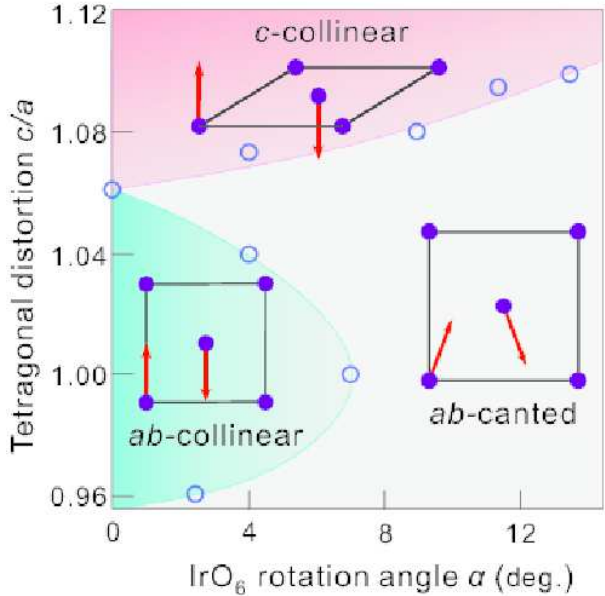}
\end{center}
\caption{(color online) \emph{Ab initio} magnetic phase diagram of Sr$_2$IrO$_4$ in the $c/a$-$\alpha$ space.
Open (blue) circles refer to points selected from the calculated data to demarcate the phase
transition boundary. The insets illustrate the specific magnetic ordering in the three regions.}
\label{fig:04}
\end{figure}

By conducting a series of additional calculations for several $c/a$ and $\alpha$ values we obtained the
\emph{ab initio} magnetic phase diagram shown in Fig. \ref{fig:04}, illustrating
the regions of stability in the $c/a$-$\alpha$ space of the three different magnetic phases
introduced  in Fig. \ref{fig:03}: $ab$-canted AFM, $ab$-collinear AFM and $c$-collinear AFM.
Our results demonstrate that structural modifications can drive two types of magnetic
transitions: (i) a spin-flop Morin-like transition, which is controlled by the
tetragonal distortion and occurs at any value of $\alpha$, and (ii) an intraplane canted to collinear
transition emerging only at small roations ($\alpha < 7^{\circ}$ and around the cubic
phase ($c/a= 1\pm0.06$).
The geometrical tunability of the magnetic ordering in Sr$_2$IrO$_4$ is made possible by the
coexistence and competition of AFM-isotropic and DM interactions, the latter being
particulary strong due to the large spin-orbit strength.

In conclusion, by combining relativistic noncollinear constrained DFT+$U$
calculations with the spin-dependent energy model, we have evaluated the isotropic and
Dzyaloshinskii-Moriya exchange parameters of Sr$_2$IrO$_4$ and built a structure/magnetism phase
diagram fully \emph{ab initio}, involving two types of structure-induced canted to collinear
magnetic transitions. The magnetic interaction parameters are strongly affected by the lattice
distortions, appealing for the direct manipulation of complex magnetic states by small structural
perturbations.

\begin{acknowledgements}
This work was supported by China Scholarship Council (CSC)-Austrian Science Fund (FWF) Scholarship Program,
by the joint FWF and Indian Department of Science and Technology (DST) project INDOX (I1490-N19), and
by the FWF-SFB ViCoM (Grant No. F41).
Computing time at the Vienna Scientific Cluster is greatly acknowledged.
\end{acknowledgements}

\end{document}